# Multiplexing Capabilities of Cold-Neutron Triple-Axis Spectrometer SIKA

Guochu Deng[a*], Peter Vorderwisch[b,c], Garry J McIntyre[a]

[a] *Australian Centre for Neutron Scattering, Australian Nuclear Science and Technology Organization, New Illawarra Road, Lucas Heights, NSW 2234, Australia*

[b] *Department of Physics, National Central University, Jhongli 32054, Taiwan*

[c] *Helmholtz-Zentrum Berlin für Materialien und Energie, Hahn-Meitner-Platz 1, Berlin D-14109, Germany*

**Abstract**

SIKA, a high-flux cold-neutron triple-axis spectrometer at the OPAL reactor at the Australian Nuclear Science and Technology Organization, is equipped with a 13-blade analyser and position-sensitive detector. This multiplexing design endows SIKA high flexibility to run in both traditional triple-axis and multiplexing analyser modes. In this study, two different multiplexing modes on SIKA are simulated using Monte-Carlo ray-tracing methods. The simulation results demonstrate SIKA's capabilities to work in these operational modes, especially, the multi-$Q$ const-$E_f$ mode. This capability was demonstrated by measuring the phonon dispersion of a Pb single-crystal sample with the multi-$Q$ const-$E_f$ mode on SIKA. Compared to the traditional and multi-analyser triple-axis spectrometers, multiplexing modes on SIKA combine the advantages of the high data-acquisition efficiency and flexibility to focus on local areas of interest in the ($Q$, $\hbar\omega$) space.

*E-mail: guochu.deng@ansto.gov.au

## 1. Introduction

A triple-axis spectrometer (TAS) is a powerful tool to investigate quasi-particle dynamics in condensed matter, such as phonon and magnon excitations. Most of TASs currently used follow a traditional design, which uses a single detector (see Fig. 1(a)) and only allows exploration of ($Q$, $\hbar\omega$) space point by point. Since only a small fraction of the scattered beam is detected at one point, this design is highly inefficient in data acquisition, especially when the ($Q$, $\hbar\omega$) space of interest is large.

In recent years, many efforts have been made to design and build new types of TAS to allow measurements at a series of points in ($Q$, $\hbar\omega$) space simultaneously (see Fig. 1(b)).[1-5] Some instruments have been built by introducing a so-called multiplexing analyser equiped with a position-sensitive detector (PSD). In such a design, the traditional analyser with an array of crystals is substituted by a series of analyser blades. Each blade on the multi-blade analyser is able to rotate independently to any specific angle, which allows detectors to collect the scattered neutrons over a large range of reciprocal space ($Q$) or energy transfer ($\hbar\omega$) at the same time. Thus, each analyser blade corresponds to a signal channel, which substantially improves the data acquisition efficiency. In order to avoid cross-talk between neighbouring signal channels, normally, radial pre-analyser and pre-detector collimators are included. RITA-1 at Risø,[2] BT7[6] and SPINS[7] at NIST are typical examples of the unconventional multiplexing analyser TAS. Another type of design to improve the data-acquisition efficiency is the multi-analyser/detector system, in which there is a series of independent analysers and detectors, arranged in the scattering plane and covering a quite large scattering angle. In this design, normally, the analyser and detector are designed for a few fixed final energies, which sacrifices the flexibility. For example, the FLATCONE option[3] for IN8, IN14 and IN20 at the Institut Laue-Langevin (ILL) consists of two sets of 31 analysers, covering a 75° scattering-angle range. Not only can two different final energies ($E_f$ =1.5 Å$^{-1}$ and 3 Å$^{-1}$) be chosen for experiments, but also the whole detector system can be tilted to collect data out of plane in momentum space. MultiFLEXX is another multi-analyser TAS at Helmholtz-Zentrum Berlin (HZB)[4]. It provides 31 analyser-detector channels covering 75° scattering angle range. Each analyser-detector channel consists of five independent analysers and detectors, which correspond to final energies of 2.5, 3.0, 3.5, 4.0 and 4.5 meV. This means that it is able to map a $Q$ space at five independent energy transfers simultaneously.

Table I. Comparison of traditional and unconventional TAS designs

| TAS type | Traditional TAS | Multiplexing analyser | Multi-analyser/detector system |
|---|---|---|---|
| Data acquisition | point by point | multichannel | Multichannel |
| Scattering angle | single | ~10-15° | Quite large (e.g. 70° for Flatcone) |
| Energy channel | multichannel | quite large | Depending on number of analysers in each channel |
| Flexibility | normal | high | low |
| Cross talk | no | depending on configuration | no |
| Examples | IN8, IN14 | RITA, SPINS, BT7, SIKA | FLATCONE, MultiFLEXX |

Both multiplexing and multi-analyser TASs improve the data-acquisition efficiency, compared to a traditional single analyser/detector TAS. However, a multiplexing TAS provides more flexibility in instrument configurations and measurements while a multi-analyser TAS covers a wider $Q$ range simultaneously. TABLE I compares the designs and advantages of the traditional and unconventional TASs.

SIKA is a high-flux RITA-type cold-neutron TAS, which is constructed at the reactor face of the OPAL reactor of the Australian Nuclear Science and Technology Organization (ANSTO).[8] The design of SIKA is similar to BT7, the double-focusing thermal-neutron TAS at NIST.[6] The secondary spectrometer of SIKA is composed of a multi-blade analyser, including 13 PG (002) blades, a linear PSD, and a separate single detector. It is able to operate in a traditional point-by-point mode by using the single detector or in different multiplexing modes with the PSD, e.g. focusing-analyser mode, $q$-dispersive modes, and energy-dispersive modes. Theoretically, the multiplexing operation modes could be realized in many different ways with various collimator-blade detector configurations. The basic idea is to extend the $Q$ or energy detecting range with one scattering configuration by driving the analyser blades to a series of specific angles, which improves the data acquisition efficiency by combining with the efficiency of the PSD.

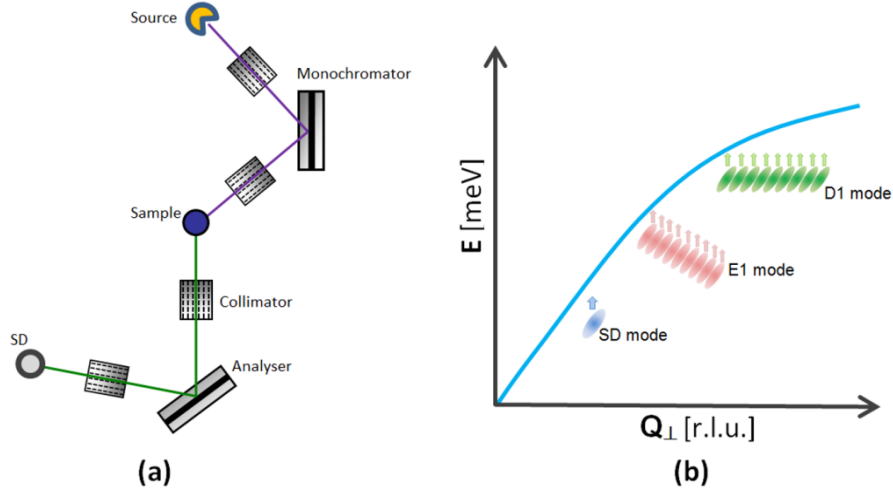

Fig. 1 (a) Traditional single-detector TAS mode; (b) Comparison of scans by using the single-detector mode, multiplexing E1 mode, and multiplexing D1 mode; The solid curve shows a dispersion curve to be measured. The blue ellipsoid denotes single-detector resolution, the pink ellipsoids denote resolutions of E1-mode channels, and the green ellipsoids denote resolutions of D1-mode channels. Arrows shows the scan directions.

In this study, we simulate the two multiplexing operation modes on SIKA by using the Monte-Carlo ray-tracing technique, including the energy-dispersive flat-analyser mode (the so-called E1 mode in Ref [9]) and the multi-$Q$ const-$E$ mode (the so-called D1 mode in Ref[9]) (see Fig. 1(b)). The energy resolution of each signal channel of these modes is calculated and presented. For the D1 mode, not only has the simulation been conducted, but also an experiment was conducted on a Pb single crystal to measure its phonon dispersion with the configuration of this mode. Analysis of the real data has been conducted and presented. These simulations and experiments demonstrate the capability of multiplexing modes of SIKA, and provide a profound understanding to multiplexing design and guidelines for conducting a successful multiplexing experiment on SIKA.

## 2. Simulations and Experiments

The simulations were conducted by using the Monte Carlo ray-tracing package RESTRAX by J. Šaroun[10]. The simulation has been done on a dummy Pb single-crystal sample with a cubic structure with lattice parameter $a$ = 4.98Å. The geometric configuration for the neutron source, guide, monochromators, analyser and collimators are based on the current instrument geometry of SIKA with some minor simplifications. The pre-monochromator, pre-sample, pre-analyser, and pre-detector collimators are all 60'. The Pb single-crystal sample was modelled with the two perpendicular scattering vectors, (1 0 0) and (0 1 1) in the equatorial plane. The simulation was performed near the sample reciprocal-lattice vector $Q$ (2 0 0), where the phonon structure factor is strong for a real experiment. For the E1 mode, we chose the most-frequently-used final energy 5meV for the simulation. The energy resolution was simulated near the elastic line. Each analyser blade was considered as an independent analyser to form a signal channel for the simulation. The difference between successive analyser blades is the angle between the scattered neutron beam and the analyser blades, which results in a series of final energies. For the D1 mode, the final energies of all blades are set to 5meV or 8.07meV for the two different high-order-contamination filter options. The former final energy is the most frequently used one for the Be filter. The latter final energy is one good choice when using a PG filter on SIKA.

Table 2. Instrument configurations for simulating the E1 and D1 modes on SIKA

| Mode | Monochromator mosaicity (PG200) | Analyser mosaicity (PG200) | Collimator divergence | $K_i$(Å$^{-1}$) | $K_f$(Å$^{-1}$) | Q(r.l.u.) |
|---|---|---|---|---|---|---|
| E1 mode | 35' | 35' | All 60' | 1.5534 | 1.5534 | ~(2, 0, 0) |
| D1 mode ($E_f$ = 5.00meV) | 35' | 35' | All 60' | 1.5534 | 1.5534 | ~(2, 0, 0) |
| D1 mode ($E_f$ = 8.07meV) | 35' | 35' | All 60' | 1.9735 | 1.9735 | ~(2, 0, 0) |

A neutron experiment was conducted on a Pb single-crystal sample with the multi-$Q$ constant-$E_f$ mode at the final energy $E_f$ = 8.07meV. Before starting the experiment with the sample, the monochromator and analyzer angles of SIKA were calibrated with the central analyzer blade. The counting efficiency of each wire in the PSD was measured and calibrated by using a polycrystalline vanadium rod as a sample. Similarly, calibration of each blade angle for the same final energy was conducted by using the same vanadium sample and rocking each blade to maximize the intensity on the PSD. The Pb single crystal was placed in air and aligned by using two independent Bragg reflections (200) and (022), which were used to build the orientation (UB) matrix.[11, 12] The longitudinal and transverse phonon dispersion curves were measured near the zone center $Q$ (200). The data were normalized by using the vanadium scans mentioned above. The normalized data were treated by using a Python code to generate a 3D dataset in the $Q_H$-$Q_{KL}$-E space.

## 3. Simulation results and discussions

### 3.1 Simulation of energy-dispersive flat-analyzer mode

The instrumental configuration of the energy-dispersive E1 mode is shown in Fig. 2. As can be seen in Fig.2 (a), the analyzer blades remain flat just like a traditional unfocused analyzer. The most important difference between the traditional mode and the E1 operation mode is the radial collimator between sample and analyser, which separates the scattered neutron signals from the sample into 13 channels. Going through each analyser blade, the neutron beam in each channel reaches the PSD and generates signals at different areas of the PSD. Since the incident angles on the analyser blades are different, the corresponding final energy $E_f$ at each blade is different too. This is demonstrated in Fig. 2(b) by using different colours for the 13 channels. Fig. 2 (c) displays the scattering triangles of different blades in this operational mode. It is clear that the $K_f$ angle and modulus change for different blades.

In the E1 mode, the divergence angle, $\delta_i$, which is the angle between beam to the $i$th blade and the central blade from the sample, is given by the following equation:

$$\delta_i = \arctan\left( d_i * \sin(\frac{A_2}{2}) / \left( L_{SA} - d_i * \cos(\frac{A_2}{2}) \right) \right) \qquad \text{Eq(1)}$$

where $L_{SA}$ is the distance between sample and analyser, $d_i$ the distance of the $i$th blade from the centre, and $A_2$ is the take-off angle of the analyser. The diffraction angle for the $i$th analyser blade is $A_2/2 + \delta_i$, which also defines $K_f$ of the outcoming beam for this blade. The angle between $K_i$ and $K_f$ is $S_2 + \delta_i$, where $S_2$ is the sample take-off angle. With all these geometric relationships and the diffraction law, we are able to calculate the configuration of each blade for the Monte-Carlo simulation.

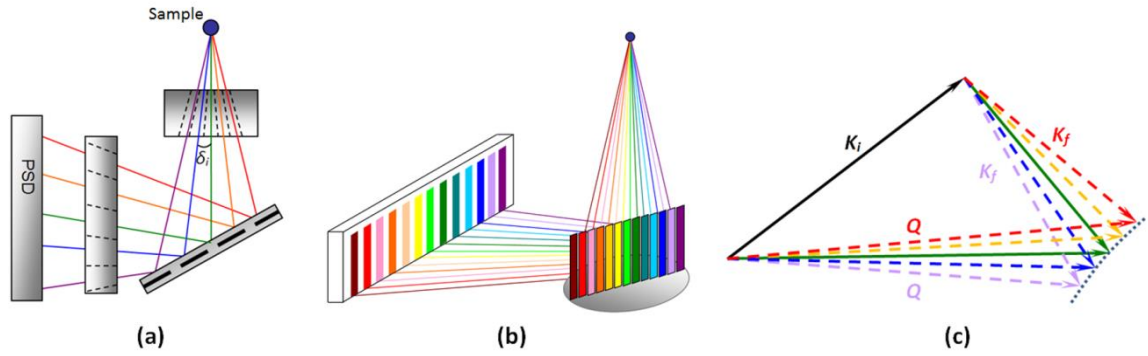

Fig. 2 (a) Configuration of the analyzer and detector system for the energy-dispersive flat-analyser mode (namely, the E1 mode); (b) Schematic of the scattering and data acquisition channels through all analyzer blades in the E1 mode; colors on the PSD indicate the different $E_f$ of each channel; (c) Scattering triangle of the E1 mode, where the final wavevector $K_f$ of each neutron scattering channel varies due to different angles between sample and each analyzer blade. A series of data with different $Q$ and $E_f$ are collected simultaneously.

From the discussion above, we could consider each analyser blade as an independent analyser to calculate its resolution ellipsoid. Actually, these analyser channels look like a series of traditional TAS measurement operating at the same $E_i$ but 13 slightly different $E_f$, which is schematically shown by the pink ellipsoids in Fig. 1(b). For simplicity, we adopted a post-sample Soller collimation (60') for the simulation since each blade is considered as an independent signal channel.

All the resolution ellipsoids of the signal channels demonstrated in Fig. 2(b) are simulated. The results from the odd-number blades are plotted in Fig. 3(a) and (b) in the $Q_{//}$-E and $Q_\perp$-E spaces, respectively. This figure shows that the final energies of all those blades cover a range from -0.6meV to 0.5meV compared to the final energy of the central blade. The momentum shifts along the $Q_{//}$ direction is smaller than the shifts along the $Q_\perp$ direction. The resolution ellipsoids in the $Q_{//}$-E spaces heavily overlap each other while those resolution ellipsoids in the $Q_\perp$-E space are well separated. This means that this mode is more efficient to map the $Q_\perp$-E space than the $Q_{//}$-E space since it covers a larger area of interest in the $Q_\perp$-E space in one single scan. Since the $K_f$ of each signal channel are slightly different, the energy profiles of the resolution ellipsoids are different too. The full width at half maxima (FWHM) obtained by fitting to these profile curves is shown in TABLE 3. The resolutions are slightly improved with the increase of the scattering angle, which is consistent with the results in Fig. 3(a) and (b). The FWHM of the central blade is 0.128meV. Considering the whole energy range (~1meV) covered by all these channels, we are able to collect data over a wide energy range with high resolution in one single scan in this operational mode. This simulation demonstrates the efficiency and advantages of the multiplexing capability on SIKA.

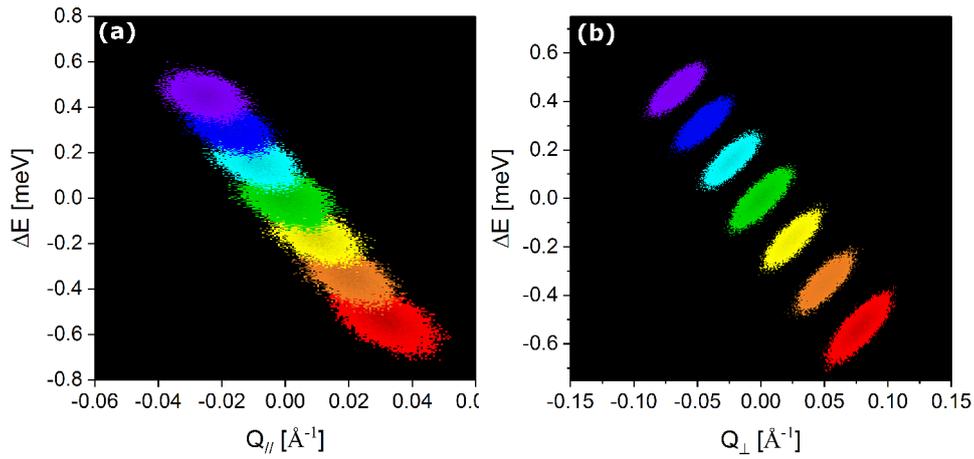

Fig. 3 Resolution ellipsoids of all the analyzer channels in the (a) $Q_\perp$-E and (b) $Q_{//}$-E spaces for the E1 mode where ΔE of the central analyzer blade is zero and $E_f$=5meV.

Table 3 Energy resolutions at different blade channels of the E1 mode.

| Blade No. | 1 | 3 | 5 | 7 | 9 | 11 | 13 |
|---|---|---|---|---|---|---|---|
| FWHM (meV) | 0.14300 | 0.13996 | 0.13127 | 0.12789 | 0.11997 | 0.11351 | 0.11065 |
| Err (meV) | 7.17E-04 | 7.83E-04 | 0.0011 | 9.11E-04 | 8.56E-04 | 9.22E-04 | 8.23E-04 |

## 3.2 Simulation of the multi-$Q$ constant-$E_f$ mode

Fig. 4 shows the schematic configuration of the multi-Q constant-$E_f$ mode. In this mode, each analyzer blade rotates to a certain angle in order to fulfill the condition that all the final energies of these blades are the same (see Fig. 4(a)). Fig. 4(b) shows the total 13 signal channels corresponding to the 13 blades. Different from Fig. 2(b) for the energy-dispersive flat-analyser mode, signals on detectors in this mode have the same energy, which is shown by one single color for all the detectors. Since the incident neutron to the sample is the same for any analyzer blade, the incident energy $E_i$ (or wave vector $K_i$) is a constant. The final energy $E_f$ is fixed to be the same value on each analyzer blade, which means the modulus of $K_f$ is a constant, too. As shown by the scattering triangle in Fig. 4(c), thus, $K_f$ of each signal channel in this operational mode is the radius of a circle. In the scattering triangle, only $Q$ and the scattering angle between $K_i$ and $K_f$ change for different blades. Comparing with the E1 mode, where $K_f$ modulus and angle, and $Q$ change at the same time, this operational mode has an advantage that the data are more straightforward and easier to be combined. The difference between the E1 and D1 operational modes are presented in the schematic of Fig. 1(b). The green ellipsoids of all D1 channels have the same energy transfer while the pink ones in all E1 channels have different $E_f$.

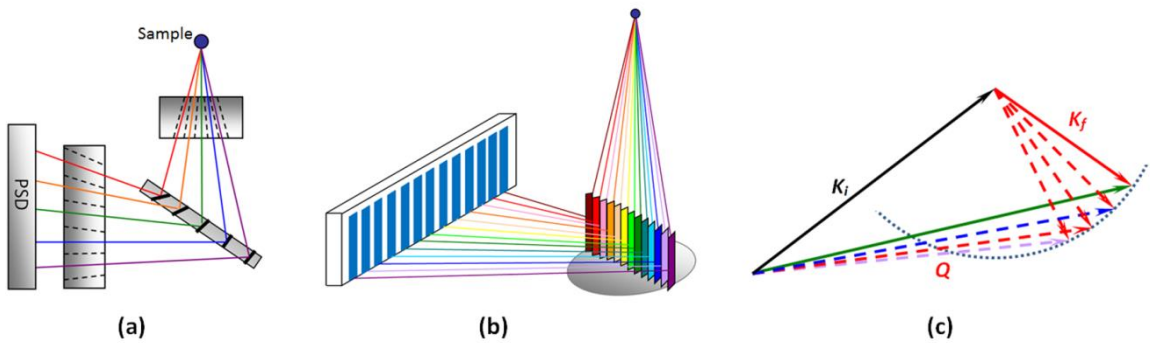

Fig. 4 (a) Configuration of the analyzer and detector for the multi-Q constant-$E_f$ mode (the so-called D1 mode); (b) Schematics of neutron path from the sample through each analyzer blade to the PSD in the D1 mode; (c) Scattering triangle of the multi-$Q$ constant-$E_f$ mode, where the final energy is constant while $Q$ varies from one analyzer blade channel to the other, allowing acquisition of data in several $Q$ channels simultaneously.

Fig. 5 demonstrates the Monte-Carlo ray-tracing instrument-resolution results of the central blade near the elastic line for the configurations of $E_f$ = 5 meV and 8.07 meV. As mentioned above, the $E_f$ for all the blades are the same. Thus, their resolution ellipsoids are quite similar and it is not necessary to present the results of each individual blade. From this plot, it is obvious that the resolution ellipsoids along the $Q\perp$ and $Q_{//}$ directions at $E_f$ = 5.0meV is much smaller than those simulated for the configuration of $E_f$ = 8.07 meV. The direct comparison of the energy profiles of these two configurations is displayed in Fig. 6. The FWHM of $E_f$ = 5 meV is about 0.125 meV while the FWHM of $E_f$=8.07 meV is about 0.28meV. These simulations give us a guideline to choose a proper configuration for a real experiment with the multiplexing analyser on SIKA.

Another important feature of this operational mode is the $Q$ range covered by the analyser blade channels, which is directly related to the subtended angle by all the blades referring to the sample position. The overall subtended angle of the blades depends on the final energy $E_f$, the chosen analyser stage angle, the distance from the sample, and the distance between neighbour blades. The analyser stage angle can be chosen in a quite flexible way. The only condition which should be fulfilled is that any analyser blade should not be shaded by any other blades on both the incoming and outcoming beam paths. Thus, an angle around of a half of the scattering angle of $E_f$ is quite a good choice to allow a wide coverage without shadowing effects. We are able to see this problem in the real experiment which will be discussed in the next section.

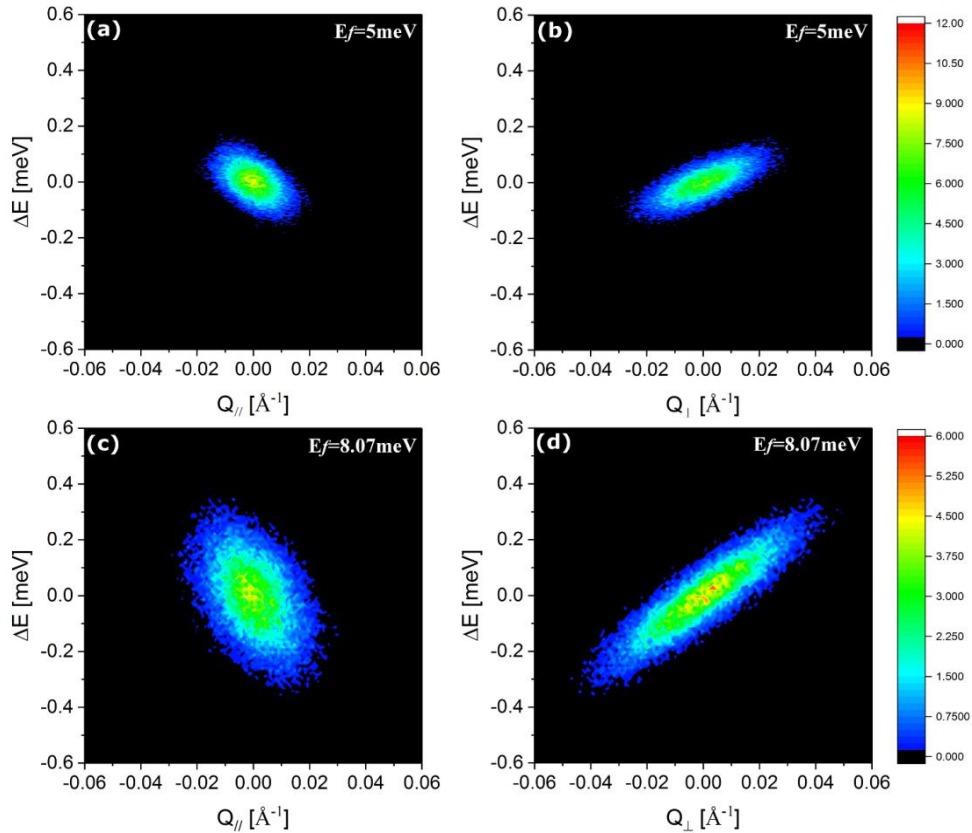

Fig. 5 The resolution ellipsoids of the central analyzer channel in the (a) $Q_\perp$-E and (b) $Q_{//}$-E spaces for the D1 mode when ΔE of the central analyzer blade is zero and $E_f$= 5 meV, and the resolution ellipsoids of the central analyzer channel in the (c) $Q_\perp$-E and (d) $Q_{//}$-E spaces for the D1 mode when ΔE of the central analyzer blade is zero and $E_f$ = 8.07 meV.

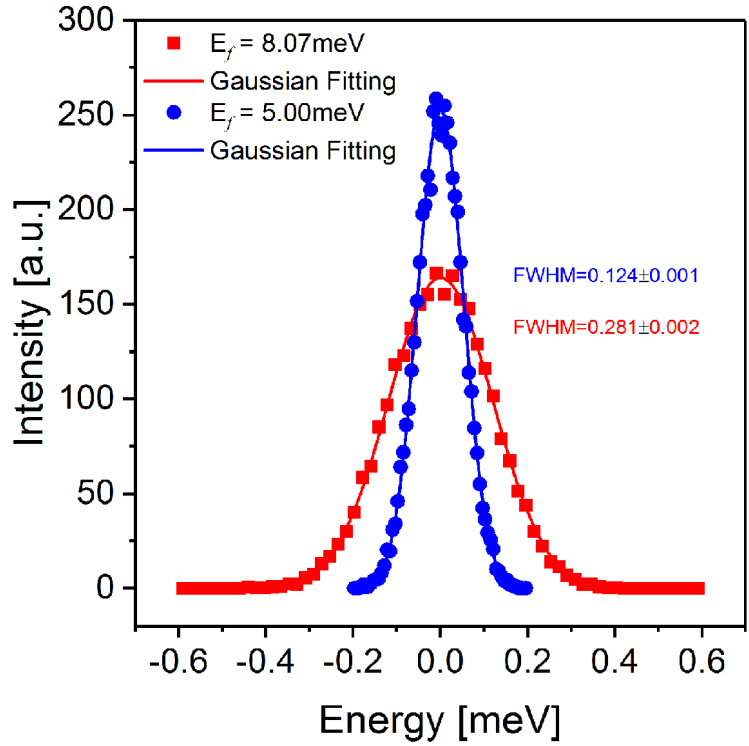

Fig. 6 The energy resolutions of the central blade channel in the D1 mode near the elastic line with $E_f$ = 5 meV and $E_f$ = 8.07 meV.

## 4. Inelastic neutron experiment with the multi-$Q$ const-$E_f$ mode

In order to carry out an experiment with the multi-$Q$ const-$E_f$ operational mode, we should take several effects into account beforehand. First of all, the SIKA PSD has 48 counting wires, which need a calibration for the counting efficiency. Such a calibration can be done by either counting a standard vanadium sample at a static position or scan all the PSD wires through a Bragg peak of a standard nickel powder sample. In the current experiment, we use the latter to calibrate the counting efficiency of each wire of the SIKA PSD. Secondly, each analyser blade should be individually calibrated for the same final energy $E_f$. This could be done by rocking each analyser blade with a standard vanadium rod at the sample position, maximizing the intensity on the PSD.

Finally, the counting efficiency of one each analyser blade channel could be different due to the geometric effects since the neutron flight path of each blade channel is slightly different. The neutron optics such as collimators could be not perfectly the same for each channel, too. Thus, a second efficiency calibration should be considered for the purpose of normalizing the data after the experiment. In order to do so, a standard vanadium rod was introduced to count at the elastic line of the current operational mode. The data are plotted in Fig. 7. It is worthwhile to mention that we only used 9 blades for the current commissioning experiments due to some geometric limitations, introduced by the sample position not being exactly at the geometric centre of the radical collimator after the sample. However, it is still possible to use 11 or even 13 blades in this mode by improving the geometric conditions in the future since we still have a few PSD wires presently not in use at both ends. As can be seen, the signal from each analyser blade mainly goes to three neighbour wires and is well separated from the signals from other blades. The separation of the signal channels is very important for the current operational mode, otherwise, the crosstalk of neighbour channels could ruin the real signals from the sample. The crosstalk for a similar operational mode on RITA-II was discussed by Bahl et al. [13]

In the current experiment, we collected the data from the Pb single crystal along the longitudinal and transverse directions at room temperature. In order to test the crosstalk of each analyser channels, we carried out the longitudinal scans without and with the pre-PSD collimator, which are shown in Fig. 8(a) and (b), respectively. As clearly shown in these two figures, the signals on the PSD without the collimator are much noisier than the data with the collimator, especially at the low-$q$ range. This indicates that the crosstalk is a critical issue when operating a multiplexing mode.

Fig.7 (b) and (c) show the Pb-crystal phonon-dispersion data collected along longitudinal and transverse directions, using this multi-$Q$ mode. These data have been normalized by considering the geometric effect mentioned above. The data have also been converted from the real space into the reciprocal space according to the orientation (UB) matrix[11, 12] built during the experimental setup. As can be seen, the 9 channels in the current setup cover about (-0.06, 0.06) r.l.u. along the $Q_H$ directions and about (-0.032, 0.032) r.l.u. along $Q_{KL}$ directions along the longitudinal direction. Thus, only five scans give us a full map of the longitudinal phonon branch of our Pb single-crystal with 45 individual $Q$ points. However, due to the geometric setup, these points are not completely distributed in the $Q_{KL}$=0 plane, but with a small $Q_{KL}$ offset. The $Q_{KL}$ component is limited in (-0.035, 0.035) r.l.u., and even smaller range at high-energy transfer. Along the transverse direction, the 9 channels covered a slightly smaller $Q_{KL}$ range in one scan, especially at the higher-energy regime. This can be clearly seen from the gap between the two scans in Fig. 8(c). At the Pb phonon signal level, one scan covers about (-0.035, 0.035) r.l.u. range in $Q_{KL}$ while it covers (-0.057, 0.057) r.l.u. along the $Q_H$ direction. From these results, we found that the multi-$Q$ const $E_f$ mode is more efficient in measuring longitudinal modes than measuring transverse modes. The perpendicular component is normally quite small, which may be ignored for an experiment not requiring high accuracy.

It is interesting to compare the current operation mode with the traditional single-detector mode on SIKA. Normally, a Soller collimator of the same divergence rather than a radial collimator to be used on the pre-analyser position, only the central blade (~20mm in width) will be illuminated by the scattered beam from a sample of the size about 20mm in diameter. Thus, the wide area of the analyser is not really useful to increase the counting efficiency, especially when we use cold neutrons such as $E_f$ = 5meV or below. When we use a radial collimator at the pre-analyser position in the multi-Q const-$E_f$ mode, each channel has similar count rate as the traditional single detector mode assuming the same sample size. Thus, the counting efficiency is 9 times faster (if we use 9 channels like in the current experiment) than the traditional single-detector system.

We should also draw our attention to the comparison between the multiplexing system in this study and the multi-analyser detector system such as Flatcone and MultiFLEXX. Actually, Flatcone and MultiFLEXX are designed for the multi-$Q$ const-$E_f$ mode as we discussed above. Both of them have a series of analyser channels with the const-$E_f$ setup. The main difference is that the analyser/detector channels of Flatcone[3] and MultiFLEXX[4] have a scattering angle step of 2.5°, much larger than the angle step (~0.53°) in the current experiment. If we use 2.5° for the angle between the signal channels in this experiment to treat the data, we find that the $Q$ step is much larger and one single scan will cover the $Q_H$ range about ~0.25 r.l.u. and $Q_{KL}$ range 0.39 r.l.u. Some of the data are far out of the range of interest. From this point of view, Flatcone and MultiFLEXX are more suitable to do an overview survey experiment in the whole momentum energy space, just like a time-of-flight spectrometer, while the current multiplexing mode on SIKA is more suitable to conduct local scans near the area of interest with more detail. Therefore, both multiplexing and multi-analyser techniques substantially improve the data acquisition efficiency, and are suitable for scans of different purposes. We note that Flatcone and MultiFLEXX cover 31 channels, much more than 9 channels we used for comparison above, over a much larger $Q$ range than the range shown above.[3,4]

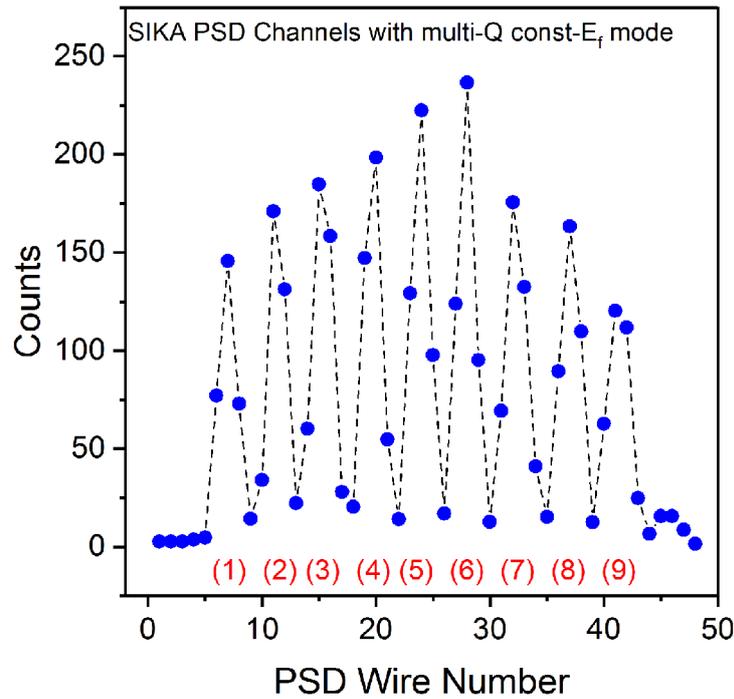

Fig. 7 The signals collected on the PSD wires with a standard vanadium sample rod at the sample position with the multi-Q const-$E_f$ mode configuration. The peak intensity was used to normalize the intensity of each channel in this mode.

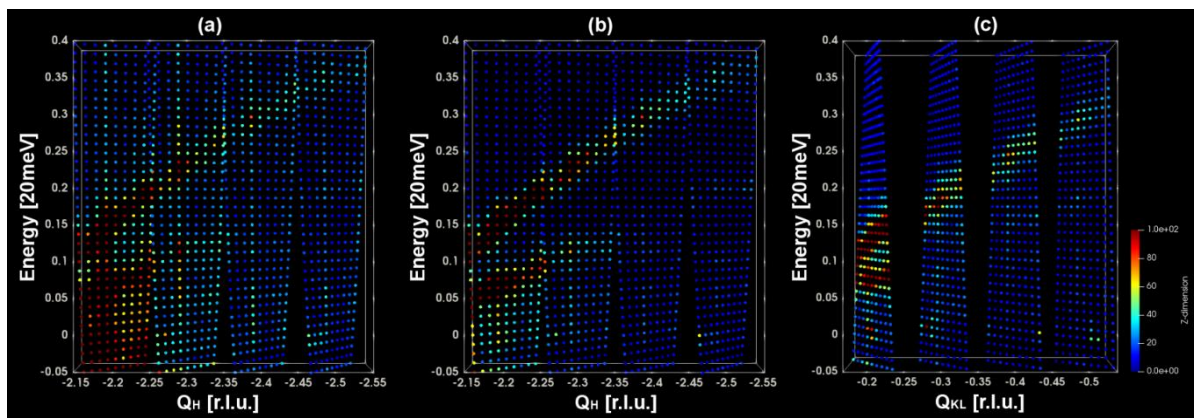

Fig. 8 (a) The phonon excitation dispersions along the longitudinal ($Q_H$) direction without the pre-detector collimator (a), along the longitudinal ($Q_H$) direction with the pre-detector collimator (b), and along the transverse ($Q_{KL}$) direction with pre-detector collimator (c) measured from the Pb single-crystal sample with the multi-$Q$ const-$E_f$ (D1) mode; The Q directions into the page are $Q_{KL}$ (a), $Q_{KL}$ (b) and $Q_H$ (c). The intensity signals have been normalized according to the detector efficiency of each wire from the standard vanadium-rod scans. Both longitudinal and transverse results shown here agree well with the phonon dispersion of Pb single crystal previously reported by Brockhouse et al.[14]

## 5 Conclusion

In summary, SIKA is a cold-neutron triple-axis spectrometer with the multiplexing analyser design. We discuss the two different multiplexing modes on SIKA: the energy-dispersive flat-analyser mode and the multi-$Q$ const-$E_f$ mode. Based on the exact SIKA geometry, Monte-Carlo ray-tracing simulations have been done for both these modes and demonstrate the feasibility of these modes on SIKA. Phonon dispersion of a Pb single-crystal sample was measured on SIKA by using the multi-$Q$ const-$E_f$ mode. After normalization and treatment, the dispersion curve of Pb was presented in a 3D momentum-energy space. This experiment demonstrates the capability of SIKA in the multiplexing operation modes. Comparison between the multiplexing mode on SIKA and the multi-analyser triple axis instruments are discussed.

**Acknowledgments**

The SIKA project was financially supported by the grants from the Ministry of Science and Technology of Taiwan (MOST, Grant No.: NSC 94-2739-M008-001 and 100-2112-M-213-006-MY3). The operation of SIKA is currently funded by the National Synchrotron Radiation Research Centre (NSRRC), Taiwan. G. D cordially thank Prof. Wen-Hsien Li (National Central University) for the leadership of SIKA design and construction, Dr. Shih-Chun Chung (NSRRC) for leading the current Taiwanese neutron team, Dr. Chun-Ming Wu (NSRRC) and Dr. Shin-ichiro Yano (NSRRC) for participating the SIKA PSD commissioning, Jen-Chih Peng (NSRRC) for the SIKA PSD data acquisition support, Eno Imamovic for the SIKA design support, and Dr. Jan Šaroun (Nuclear Physics Institute, ASCR, Řež, Czech Republic) for his help with RESTRAX.

**References**

1   K. N. Clausen, D. F. McMorrow, K. Lefmann, G. Aeppli, T. E. Mason, A. Schröder, M. Issikii, M. Nohara, and H. Takagi, Physica B: Condensed Matter **241-243**, 50 (1997).

2   T. E. Mason, T. E. Mason, K. N. Clausen, G. Aeppli, D. F. McMorrow, J. K. Kjems, and G. Aeppli, Canadian Journal of Physics **73**, 697 (1995).

3   M. Kempa, B. Janousova, J. Saroun, P. Flores, M. Boehm, F. Demmel, and J. Kulda, Physica B: Condensed Matter **385-386**, 1080 (2006).

4   F. Groitl, R. Toft-Petersen, D. L. Quintero-Castro, S. Meng, Z. Lu, Z. Huesges, M. D. Le, S. Alimov, T. Wilpert, K. Kiefer, S. Gerischer, A. Bertin, and K. Habicht, Scientific Reports **7**, 13637 (2017).

5   F. Groitl, D. Graf, J. O. Birk, M. Markó, M. Bartkowiak, U. Filges, C. Niedermayer, C. Rüegg, and H. M. Rønnow, Review of Scientific Instruments **87**, 035109 (2016).

6   J. W. Lynn, Y. Chen, S. Chang, Y. Zhao, S. Chi, W. Ratcliff, B. G. Ueland, and R. W. Erwin, Journal of Research of the National Institute of Standards and Technology **117**, 61 (2012).

7   S. F. Trevino, J Res Natl Inst Stand Technol. **98**, 59 (1993).

8   C.-M. Wu, G. Deng, J. S. Gardner, P. Vorderwisch, W.-H. Li, S. Yano, J.-C. Peng, and E. Imamovic, Journal of Instrumentation **11**, 10009 (2016).

9   K. Lefmann, D. F. McMorrow, H. Ronnow, K. Nielsen, K. N. Clausen, B. Lake, and G. Aeppli, Physica B: Condensed Matter **283**, 343 (2000).

10  J. Šaroun and J. Kulda, Physica B: Condensed Matter **234-236**, 1102 (1997).

11  W. R. Busing and H. A. Levy, Acta Cryst. **22**, 457 (1967).

12  M. D. Lumsden, J. L. Robertson, and M. Yethiraj, J Appl. Cryst. **38**, 405 (2005).

13    C. R. H. Bahl, K. Lefmann, A. B. Abrahamsen, H. M. Rønnow, F. Saxild, T. B. S. Jensen, L. Udby, N. H. Andersen, N. B. Christensen, H. S. Jakobsen, T. Larsen, P. S. Häfliger, S. Streule, and C. Niedermayer, Nuclear Instr. Methods in Phys. Res. **246**, 452 (2006).

14    B. N. Brockhouse, T. Arase, G. Caglioti, K. R. Rao, and A. D. B. Woods, Physical Review **128**, 1099 (1962).